\documentclass[letterpaper, 10 pt, conference]{ieeeconf}  % Comment this line out if you need a4paper

\IEEEoverridecommandlockouts                              % This command is only needed if 
\usepackage{graphicx} % Required for inserting images
\usepackage{amsmath}
\usepackage{amsfonts}
\usepackage{bm}
\usepackage{cancel}
\usepackage{siunitx}
\usepackage{cite}

\allowdisplaybreaks

\DeclareMathOperator{\argmin}{arg\,min}

\newtheorem{theorem}{Theorem}
\newtheorem{definition}{Definition}

\usepackage[dvipsnames]{xcolor}
\usepackage[normalem]{ulem} % for \sout in \del

\newif\ifshowcomments
\showcommentstrue             % <-- set to \showcommentsfalse to hide all comments

\ifshowcomments
  \newcommand{\cmt}[3]{\textcolor{#2}{\textbf{[#1:} #3\textbf{]}}}
  
  \newcommand{\del}[1]{\textcolor{Gray}{\sout{#1}}}
\else
  \newcommand{\cmt}[3]{}
  
  \newcommand{\del}[1]{}
\fi

\begin{document}

\title{\LARGE \bf Control Affine Hybrid Power Plant Subsystem Modeling for Supervisory Control Design}

\author{Stephen Ampleman, Himanshu Sharma, Sayak Mukherjee, Sonja Glavaski
\thanks{The work was authored by the Pacific Northwest National Laboratory (PNNL) is operated for the DOE by Battelle Memorial Institute under contract DE-AC06-76RL01830. Funding is provided by the U.S. Department of Energy Wind Energy Technologies Office for project: Path to Nationwide Deployment of Fully Coupled Wind-Based Hybrid Energy Systems. The views expressed in the article do not necessarily represent the views of the DOE or the U.S. Government. The U.S. Government retains and the publisher, by accepting the article for publication, acknowledges that the U.S. Government retains a nonexclusive, paid-up, irrevocable, worldwide license to publish or reproduce the published form of this work, or allow others to do so, for U.S. Government purposes.}
\thanks{Stephen Ampleman, Himanshu Sharma, Sayak Mukherjee and Sonja Glavaski are with the Pacific Northwest National Laboratory (PNNL).}
}

\maketitle
\thispagestyle{empty}
\pagestyle{empty}

%%%%%%%%%%%%%%%%%%%%%%%%%%%%%%%%%%%%%%%%%%%%%%%%%%%%%%%%%%%%%%%%%%%%%%%%%%%%%%%%
\begin{abstract}
Hybrid power plants (HPPs) combine multiple power generators (conventional/variable) and energy storage capabilities to support generation inadequacy and grid demands. This paper introduces a modeling and control design framework for hybrid power plants (HPPs)  consisting of a wind farm, solar plant, and battery storage. Specifically, this work adapts established modeling paradigms for wind farms, solar plants and battery models into a control affine form suitable for control design at the supervisory level. In the case of wind and battery models, generator torque and cell current control laws are developed using nonlinear control and control barrier function techniques to track a command from a supervisory control law while maintaining safe and stable operation. The utility of this modeling and control framework is illustrated through a test case using a utility demand signal for tracking, time varying wind and irradiance data, and a rule-based supervisory control law.
\end{abstract}

%%%%%%%%%%%%%%%%%%%%%%%%%%%%%%%%%%%%%%%%%%%%%%%%%%%%%%%%%%%%%%%%%%%%%%%%%%%%%%%%

\section{Introduction}

Hybrid power plants (HPPs) consist of multiple generation sources (e.g., solar PV, wind, diesel engines, etc.) alongside an energy storage system to provide efficient and reliable energy to offset consumer power demand \cite{2020NREL_OpportunitiesForResearchAndDevelopmentOfHybridPowerPlants}. These components are aggregated together in an energy management system or supervisory control architecture in order to deliver power in a safe and reliable manner \cite{2025Kaushik_ResearchChallengesAndOpportunitiesOfUtilityScaleHybridPowerPlants}. This control problem is widely recognized as one of the major pieces of development necessary for further integration of HPPs \cite{2020NREL_OpportunitiesForResearchAndDevelopmentOfHybridPowerPlants, 2025Kaushik_ResearchChallengesAndOpportunitiesOfUtilityScaleHybridPowerPlants, 2022Ammari_SizingOptimizationControlAndEnergyManagementOfHREPPs} . 
A fundamental step in progressing towards the goal of effective supervisory control of HPPs is the models underpinning the physics of these systems. These models have been well studied, and developed at varying levels of fidelity and control integration. Wind farms, for example, vary from the Jensen model \cite{1983Jensen}, which captures the fundamental properties of wind turbine wake propagation, to high fidelity Large Eddy Simulations (LES) \cite{LESGO}. Control of these systems ranges similarly as the objective changes (i.e. maximum power point tracking, frequency regulation, etc. \cite{2015Meyers_OptimalControlOfEnergyExtraction, 2018ValiPao_LargeEddySimulationStudyOfWindFarmActivePowerControlWithACoordinatedLoadDistribution}). The battery storage system (i.e. Lithium-Ion) and solar PV plants also exhibit diverse levels of fidelity in modeling and control approaches \cite{Doyle_1993, 2015Plett_BMS_Vol1, 2008Deshmukh_ModelingOfHybridRenewableEnergySystems, 2006DeSoto_ImprovementAndValidationOfAModelForPhotovoltaicArrayPerformance}. Supervisory level control often relies on these models to underpin the algorithm design \cite{2019Kaushik_DynamicModellingOfWindSolarStorageBasedHybridPowerPlant} \cite{2023Pombo_AComputationallyEfficientFormulationForAFlexibiltiyEnabling}, however, there currently exists a gap in analytical models that are suitable for control design that ensure sub-system level operational safety. There has been work in creating high fidelity hybrid power plant simulation platforms for rigorous physics based testing \cite{NREL_HERCULES_2025} or siting and market integration \cite{2024_HyDesign}, but these platforms lack a modeling framework that allows for supervisory control design.
This work's contributions are thus the following:
\begin{enumerate}
    \item An aggregation of physics-based control affine oriented modeling paradigms for an HPP based on wind, solar PV, and Li-Ion battery storage.
    \item Component-wise control laws that capture physical constraints of the system, with emphasis on wind farm and battery storage that make it a suitable test platform and basis for supervisory control law development.
\end{enumerate}

This paper is structured as follows. In Section \ref{sec:Preliminaries}, preliminaries for the component-wise control laws are discussed. Next in Section \ref{sec:HPP_MandC} the individual models of the HPPs are introduced, along with their respective control laws. Finally, in Section \ref{sec:simulation}, a supervisory control law from \cite{NREL_WHOC_2025} is integrated and a test case is demonstrated. The paper is concluded by recognizing further advancements needed in supervisory control to overcome challenges with respect to set-point tracking and simultaneous component-wise and system level constraints.

\section{Preliminaries} \label{sec:Preliminaries}

\begin{definition}[\cite{SONTAG1995351}] \label{def:ISS}
    Given a system
    \begin{equation} \label{eq:ISS_example_system}
        \dot{x} = f(x,u)
    \end{equation}
    then a smooth function $V: \mathbb{R}^{n}\to\mathbb{R}_{\ge0}$ is called an \textit{ISS-Lyapunov function} for (\ref{eq:ISS_example_system}) if $\exists$ class-$\mathcal{K}_{\infty}$ functions $\alpha_{1}$, $\alpha_{2}$ and class-$\mathcal{K}$ functions $\alpha_{3}$ and $\chi$ such that
    \begin{equation}
        \alpha_{1}(|y|)\le V(y) \le \alpha_{2}(|y|)
    \end{equation}
    for any $y \in \mathbb{R}^{n}$ and
    \begin{equation}
        \dot{V}=\nabla V f(y,\mu)\le -\alpha_{3}(|y|)
    \end{equation}
    for any $y\in\mathbb{R}^{n}$ and any $\mu\in\mathbb{R}^{m}$ such that $|z|\ge\chi(|\mu|)$
\end{definition}

\begin{definition}[\cite{2019Ames_ControlBarrierFunctions_TheoryAndApplications}\cite{2017Ames_ControlBarrierFunctionBasedQuadraticProgramsForSafetyCriticalSystems}]
    Consider a control affine system of the form
    \begin{equation} \label{eq:CBF_example_system}
        \dot{x} = f(x) + g(x)u
    \end{equation}
    where $f$ and $g$ are considered locally Lipshitz and a set $\mathcal{C}$ such that 
    \begin{equation}
        \mathcal{C}=\left\{x\in\mathbb{R}^{n}:h(x) \ge 0\right\}
    \end{equation}
    and $h:\mathbb{R}^{n}\to\mathbb{R}$ is a continuously differentiable function. The function $h$ is called a \textit{zeroing control barrier function (ZCBF)} defined on $\mathcal{D}$ with $\mathcal{C}\subseteq\mathcal{D}\subset\mathbb{R}^{n}$ if $\exists$ an extended class-$\mathcal{K}$ function $\alpha:\mathbb{R}^{n}\to\mathbb{R}$
    \begin{equation} \label{eq:ZCBF_def}
        \sup_{u\in U} \left[L_{f}h(x) + L_{g}h(x)u + \alpha(h(x)) \right] \ge 0, \ \forall x \in \mathcal{D}
    \end{equation}
    then given the ZCBF $h$ and for all $x\in\mathcal{D}$, 
    \begin{equation}
        K(x) = \left\{u\in U: L_{f}h(x)+L_{g}h(x)u + \alpha(h(x))\ge 0\right\}
    \end{equation}
    where $L_{f}$ and $L_{g}$ are defined as the Lie derivatives along $f$ and $g$, respectively.
\end{definition}
\begin{theorem}[\cite{2019Ames_ControlBarrierFunctions_TheoryAndApplications}\cite{2017Ames_ControlBarrierFunctionBasedQuadraticProgramsForSafetyCriticalSystems}]
    \textit{Let $\mathcal{C}\in\mathbb{R}^{n}$ be a set defined as the super-level set of a continuously differentiable function $h: \mathcal{D}\subset\mathbb{R}^{n}\to\mathbb{R}$. If $h$ is a ZCBF on $\mathcal{D}$ then any Lipshitz continuous controller $u:\mathcal{D}\to\mathcal{U}$ such that $u(x)\in K(x)$ will render the set $\mathcal{C}$ forward invariant such that for every $x_{0}\in\mathcal{C}$, $x(t)\in\mathcal{C}$ for all $t\in\left(t_{0},\tau_{max}\right]$}.
\end{theorem}
\begin{definition}[\cite{2022Wei_HighOrderControlBarrierFunctions}] \label{def:HOCBF}
    Let $m$ be the relative degree of the function $h(x):\mathbb{R}^{n}\to\mathbb{R}$, then a sequence of functions $\psi_{i}, i \in \left\{1,..,m\right\}$ can be defined as
    \begin{equation}
        \psi_{i}(x) = \dot{\psi}_{i-1}(x) + \alpha_{i}(\psi_{i-1}(x))
    \end{equation}
    where $\psi_{0} = h(x)$. Then a sequence of sets $C_{i}, i \in \left\{1,..,m\right\}$ can be defined as
    \begin{equation}
        C_{i} = \left\{x\in\mathbb{R}^{n}:\psi_{i-1}(x)\ge 0\right\}, i \in \left\{1,..,m\right\}.
    \end{equation}
    Then $h(x)$ is a \textit{high-order control barrier function (HOCBF)} if $\exists$ differentiable class-$\mathcal{K}$ functions $\alpha_{i}, i \in \left\{1,..,m\right\}$ s.t.
    \begin{align}
        \sup_{u\in U} [L^{m}_{f}h(x) + L_{g}L^{m-1}_{f}h(x)u + \frac{\partial^{m}h(x)}{\partial{t}^{m}} \nonumber \\
         + O(h(x) + \alpha_{m}(\psi_{m-1}(x)) ] \ge 0
    \end{align}
    Given that $h(x)$ is an HOCBF,
    \begin{align}
        K(x) = \{u \in U: L^{m}_{f}h(x) + L_{g}L^{m-1}_{f}h(x)u + \frac{\partial^{m}h(x)}{\partial{t}^{m}}  \nonumber \\
         + O(h(x) + \alpha_{m}(\psi_{m-1}(x))\ge 0 \} 
    \end{align}
\end{definition}
\begin{theorem}[\cite{2022Wei_HighOrderControlBarrierFunctions}] \label{thm:HOCBF}
    Given an HOCBF $h(x)$ with the sets $C_{i}, \ i \in \left\{1,..,m\right\}$, if $x_{0} \in C_{1}\cap,...,\cap C_{m}$, then any Lipshitz continuous controller $u(x)\in K(x)$ renders the set $C_{1}\cap,...,\cap C_{m}$ forward invariant for the dynamics in (\ref{eq:CBF_example_system}).
\end{theorem}

\section{HPP Modeling and Control} \label{sec:HPP_MandC}
The HPP system considered here comprises a wind farm, a solar plant and a Lithium-Ion battery storage component. These three components are decoupled in absence of a supervisory control law, in the sense that their dynamics are not interdependent. Each component is designed such that, at the component level, the closed loop dynamics are control affine with respect to the power set point commands. This is done intentionally so that the HPP in its entirety is control affine and thus amenable to a number of supervisory control strategies. In this section each component of the HPP will be introduced, in addition to a component level control law that allows for power set point tracking. This section will conclude by describing a heuristics-based supervisory control law which will be used to validate the HPP model. 
\subsection{Wind Farm} \label{sec:windfarm}

The wind farm model used here consists of $N_{T}$ number of wind turbines based on the specifications of the NREL 5MW reference turbine \cite{2007Jonkman_DynamicsModelingAndLoadsAnalysisOfAnOffshoreFloatingWindTurbine}. Each individual turbine uses the following mathematical model which consists solely of the aerodynamic component described by:
\begin{equation} \label{eq:wind_plant}
    \dot{\omega}_{r} = \frac{1}{J_{r}}\left(\frac{P}{\omega_{r}} - T_{g}\right)
\end{equation}
where $\omega_{r}$ is the turbine rotor speed, $J_{r}$ is the rotor moment of inertia, $T_{g}$ is the torque from the generator, and $P$ is the aerodynamic power which can be described by
\begin{equation} \label{eq:WindPower}
    P = \frac{1}{2}\rho A_{r} C_{P}\left(\lambda\right)||U_{\infty}||^{3}.
\end{equation}
In (\ref{eq:WindPower}), $\rho$ is the air density, $A_{r}$ is the rotor swept area, $C_{P}$ is the power coefficient which depends on the tip speed ratio, $\lambda = R_{r}\omega_{r}/||U_{\infty}||$, and $U_{\infty}$ is the upstream wind speed to the turbine. Typically, $U_{\infty}$ is a three-dimensional vector consisting of the $u$, $v$,  and $w$ components of the wind in a global frame, but in this study the incoming wind speed is assumed to be the wind speed magnitude so $U_{\infty} = ||U_{\infty}||$. 

When aggregating the entire farm, it is necessary to consider the velocity deficits imparted from turbine to turbine. In order to account for this velocity deficit, we use a squared superposition approach from \cite{2017Shapiro_ModelBasedRecedingHorizonControl}, which calculates the resultant wind field in the streamwise direction as
\begin{equation}
    U(x,t) = U_{\infty} - \left(\sum_{i\in \mathcal{G}_{j}} \delta u_{i}^{2}\right)^{1/2} \ j = 1,2,...,N_{T}
\end{equation}
The notation $i\in \mathcal{G}_{j}$ indicates that any turbine $i$ that is contained in the graph of turbine $j$ (i.e. imparts a deficit onto turbine $j$) will be summed. In this work, we assume static graphs that can be determined a-priori. A rectangular layout with sufficient spacing between columns to neglect spanwise interaction is considered. Typically for deficit models, a time delay \cite{2021Starke_NetworkBasedEstimationofWindFarmPowerandVelocityDataUnderChangingWindDirections} or advection-diffusion model \cite{2019Shapiro_AWakeModelingParadigmForWindFarmDesignAndControl} will be considered to accurately model the delay for the deficit to propagate down the column of turbines. Here, for simplicity, we consider a steady-state representation of the advection-diffusion model in \cite{2018Shapiro_ModellingYawedWindTurbineWakes} (a similar strategy is used in FLORIS \cite{NREL_FLORIS_2025}, which is widely used as a benchmarking tool in the wind energy community).
\begin{equation}
    \delta u_{i}(x,t) = U(x_{i},t)\frac{2a_{i}}{d_{w}^{2}}\left(1 + erf\left(\Delta x/R_{r}\sqrt{2}\right)\right)
\end{equation}
where $d_{w}$ is the expansion of the wake in the streamwise direction
\begin{equation}
    d_{w} = 1 + k_{w}\log\left(1 + e^{\Delta x/R}\right),
\end{equation}
$a$ is the axial induction factor
\begin{equation}
    a = \frac{1}{2}\left(1 - \sqrt{1 - C_{T,i}}\right)
\end{equation}
$C_{T,i}$ is the thrust coefficient of turbine $i$, $k_{w}$ is the wake expansion coefficient, $\Delta x$ is the streamwise distance between turbines, and $erf()$ is the error function.

When including a wind farm in a power tracking scenario, the farm and turbine must be considered collectively. A simple wind farm controller is considered, which evenly divides the requested power set point to each turbine (i.e. $P_{sp,i} = P_{sp}/N_{T} $ for $i = 1,2,...,N_{T}$). Next, the ability of each turbine to track the requested power signal must be considered, specifically the design of generator torque $T_{g}$ from Equation (\ref{eq:wind_plant}). Typically, generator torque is only used when wind speeds are contained in Region 2 \cite{2011PaoJohnson_ControlOfWindTurbines}, but here owing to the lower-dimensional wind turbine model, its use will be considered agnostic of wind speed region.

With the power tracking goal in mind, an error can be formed between a given power setpoint and the power of a turbine given by
\begin{equation}
    e = P_{i} - P_{sp,i}
\end{equation}
Then a Lyapunov function of the form $V(e) = \frac{1}{2}e^{2}$ can be chosen such that $\dot{V} = e\dot{e}$. In order to make the error dynamics globally asymptotically stable, we desire $\dot{e} = -e$ such that $\dot{V}$ is negative definite $\forall e \in \mathbb{R}$. 

Since the power references are slowly varying with respect to the dynamics, we can consider the rate to be zero (i.e. $\dot{P}_{sp,i} = 0$). Thus, the dynamics for the error then depend on the derivative of the power of each turbine,
\begin{align} \label{eq:tau_control}
    \dot{e} &= \dot{P}_{i} = \frac{1}{2}{\rho_{a}}{A_{r}}RU_{\infty}^{2}\frac{\partial{C_{p}}}{\partial{\lambda}}\dot{\omega}_{r} \nonumber \\
    \dot{e} &= \frac{1}{2}{\rho_{a}}{A_{r}}RU_{\infty}^{2}\frac{\partial{C_{p}}}{\partial{\lambda}}\frac{1}{J_{r}}\left(\frac{P}{\omega_{r}} - T_{g}\right) \nonumber \\
    T_{g} &= \frac{2eJ_{r}}{\rho A_{r}R_{r} U_{\infty}^{2} }\frac{1}{\frac{\partial{C_{p}}}{\partial{\lambda}}} + \frac{P}{\omega_{r}}.
\end{align}
where the second line substitutes the dynamics in (\ref{eq:wind_plant}) for $\dot{\omega}_{r}$ and the third line solves for the control $T_{g}$. This yields a control $T_{g}$ that resembles a feedback linearization trajectory tracking control law. We seek to eliminate the term $\frac{\partial{C_{p}}}{\partial{\lambda}}$ and replace it with a constant gain to avoid the need to compute this parameter in real time and to avoid a zero crossing that occurs in this parameter with respect to $\lambda$ when the turbine reaches its maximum power coefficient. Thus the term $1/\frac{\partial{C_{p}}}{\partial{\lambda}}$ is replaced with a constant $K>0$. This has ramifications on the selected Lyapunov function, with its derivative now taking the form
\begin{equation}
    \dot{V} = -e^{2}\left(K\frac{\partial{C_{p}}}{\partial{\lambda}}\right).
\end{equation}

\begin{figure}[h]
\centering
\includegraphics[width=0.5\textwidth]{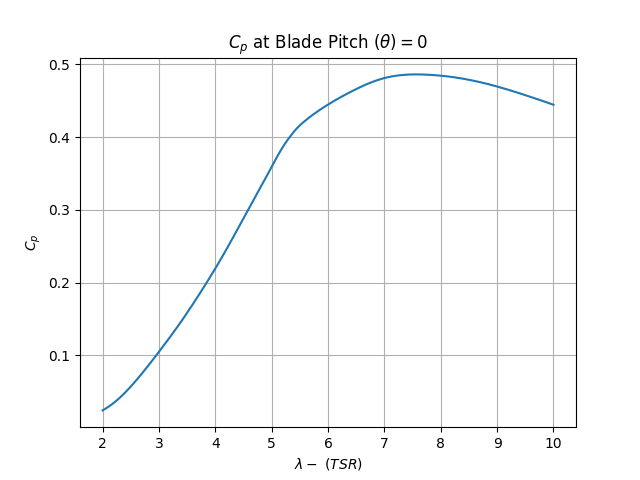}
\caption{As $\lambda$ increases beyond the maximum power point, the power curve starts to decrease, indicating a negative $\frac{\partial{C_{p}}}{\partial{\lambda}}$, which would destabilize the closed loop system with control law in (\ref{eq:tau_control})}
\label{fig:Cp_theta0}
\end{figure}

Thus the error is locally asymptotically stable when $\frac{\partial{C_{p}}}{\partial{\lambda}}>0$. Observing the curve for $C_{p}$ of the NREL 5MW turbine \cite{2007Jonkman_DynamicsModelingAndLoadsAnalysisOfAnOffshoreFloatingWindTurbine} when the collective blade pitch command is zero in Figure \ref{fig:Cp_theta0}, this is true up until the tip speed ratio where the power coefficient is maximized. Thus we can effectively limit the boundary of control for the turbine to be less than the observed maximum tip speed ratio by formulating it as a constraint of the form
\begin{equation}
    b_{\lambda}(\omega_{r}) = \lambda_{C_{p,max}} - \frac{R_{r}\omega_{r}}{U_{\infty}} \ge 0.
\end{equation}
Then $b_{\lambda}(\omega_{r})$ is a zeroing control barrier function (ZCBF) if there exists a class-$\mathcal{K}$ function $\alpha$ such that
\begin{equation} \label{eq:tau_ZCBF_1}
    \dot{b}_{\lambda}(\omega_{r}) + \alpha(b_{\lambda}(\omega_{r})) \ge 0
\end{equation}
where $\dot{b}_{\lambda}$ can be calculated by taking the Lie derivative across the dynamics in (\ref{eq:wind_plant}) such that 
\begin{align*}
    \dot{b}_{\lambda} &= \frac{\partial{b_{\lambda}}}{\partial{\omega_{r}}}\dot{\omega}_{r} \nonumber \\
    \dot{b}_{\lambda} &= \frac{-R_{r}}{U_{\infty}J_{r}}\left(\frac{P}{\omega_{r}} - T_{g}\right)
\end{align*}
Since the control appears after one derivative, this is a relative degree one ZCBF. Plugging this into (\ref{eq:tau_ZCBF_1}) and choosing a linear class-$\mathcal{K}$ function such that $\alpha(b(\omega_{r}))=c_{w}(\omega_{r})$ yields the inequality
\begin{equation} \label{eq:tau_ZCBF_ineq}
    \frac{-R_{r}}{U_{\infty}J_{r}}\left(\frac{P}{\omega_{r}} - T_{g}\right) + c_{w}(\lambda_{C_{p,max}} - \frac{R_{r}\omega_{r}}{U_{\infty}}) \ge 0
\end{equation}
Finally, setting $T_{g}^{*} = $(\ref{eq:tau_control}) allows this control problem to be cast in a control barrier function based quadratic program (CBF-QP) framework \cite{2017Ames_ControlBarrierFunctionBasedQuadraticProgramsForSafetyCriticalSystems},\cite{2014Ames_ControlBarrierFunctionBasedQPsWithApplications} such that at every time step a quadratic program is solved which computes an optimal solution to the following problem
\begin{equation} \label{eq:tau_CBFQP}
\begin{aligned}
T_{g} = \argmin_{u} \quad & \int_{0}^{T} \frac{1}{2}\left(u - T_{g}^{*}\right)^{2} dt\\
\textrm{s.t.} \quad & -\frac{R_{r}}{U_{\infty}J_{r}}u \le \frac{-R_{r}}{U_{\infty}J_{r}}\left(\frac{P}{\omega_{r}} + c_{w}\right)+ \\
& c_{w}\lambda_{C_{p,max}}
\end{aligned}
\end{equation}
In this problem, $\lambda_{C_{p,max}}$ can be designed to be slightly less than the true maximum such that any solution to the above problem will yield a $\frac{\partial{C_{p}}}{\partial{\lambda}}$ that is always greater than zero and thus $\dot{V} < 0 \ \forall e\in \mathbb{R}$ which implies the error dynamics are asymptotically stable.

\subsection{Solar Plant}

PV systems are modeled with varying degrees of fidelity, ranging from the five parameter model in \cite{2006DeSoto_ImprovementAndValidationOfAModelForPhotovoltaicArrayPerformance}, which is used to accurately predict current vs voltage curves at a given irradiance and temperature, to lower order models such as \cite{2008Deshmukh_ModelingOfHybridRenewableEnergySystems} which considers the power output given expressions for total irradiance and efficiency factors. The PV system model we consider in this work is based on \cite{2008Deshmukh_ModelingOfHybridRenewableEnergySystems} and relies on a time varying irradiance signal, the area of the solar plant, and a constant efficiency factor. Together, these parameters calculate the maximum instantaneous power of the solar plant:
\begin{equation}
    P_{max,s} = I_{T}(t)A_{s}\eta_{s}
\end{equation}
where $I_{T}(t)$ is the time-varying normal irradiance, $A_{s}$ is the total solar plant panel area, and $\eta_{s}$ is the efficiency of the solar plant. A low pass filter model is then used to illustrate the time delay for the solar cells to produce the rated power at this irradiance.
\begin{equation}
    \dot{P}_{s} = \frac{-1}{\tau}P_{s} + \frac{1}{\tau}u
\end{equation}
where $u$ in the open loop case is simply $P_{max,s}$. Given the linear nature of this model, the control input $u$ is designed via a PI control law which considers the error between the power output $P_{s}$ and power set-point $P_{s,sp}$
\begin{equation}
    e = P_{s} - P_{s,sp}
\end{equation}
\begin{equation}
    u = K_{p}e + K_{I}\int e
\end{equation}
With control law $u$, the closed loop dynamics are
\begin{equation}
    \begin{bmatrix} \dot{P}_{s} \\ e \end{bmatrix} = \begin{bmatrix}-\frac{1 - K_{p}}{\tau} & \frac{K_{i}}{\tau} \\ 1 & 0 \end{bmatrix}\begin{bmatrix} P_{s} \\ \int e \end{bmatrix} + \begin{bmatrix} \frac{K_{p}}{\tau} \\ -1 \end{bmatrix}P_{sp}
\end{equation}
The output $P_{s}$ of this model is then saturated by the maximum available power given by $P_{s,max}$.

\subsection{Battery} \label{sec:batteryMandC}
\begin{figure}[h]
\centering
\includegraphics[width=0.5\textwidth]{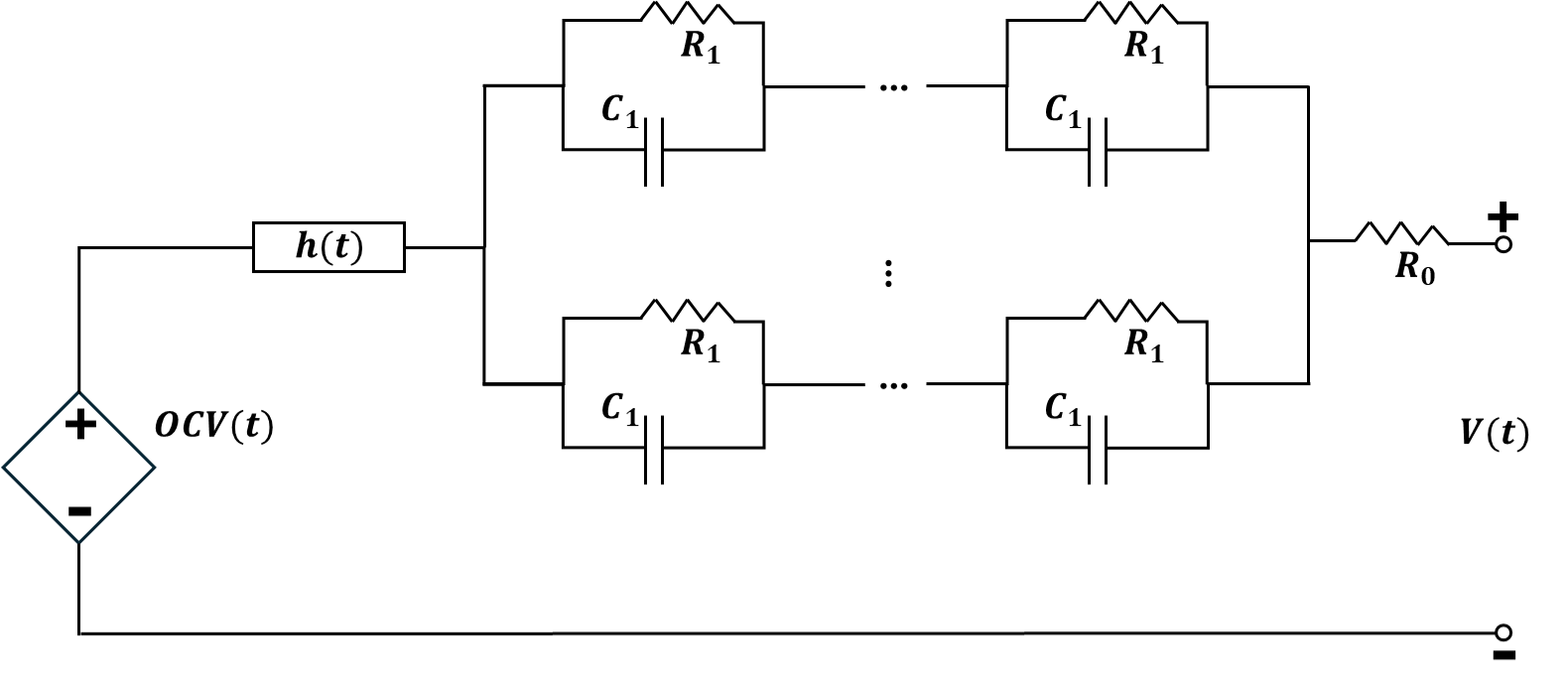}
\caption{Equivalent circuit model of a Li-Ion battery with homogeneous RC-circuit cells connected in series and parallel together with an equivalent series resistance, hysteresis voltage, and open open circuit voltage (figure inspiration from \cite{2015Plett_BMS_Vol1}, Fig. 2.12)}
\label{fig:ECM}
\end{figure}
In this work, an Equivalent Circuit Model (ECM) with single state hysteresis is considered, as shown in Figure \ref{fig:ECM}. This model takes the form
\begin{equation} \label{eq:batteryDynamics}
    \begin{bmatrix} \dot{U}_{1} \\ \dot{z} \\ \dot{h} \end{bmatrix} = 
    \begin{bmatrix}\frac{-1}{R_{1}C_{1}}U_{1} + \frac{1}{C_{1}}I_{c} \\ \frac{\eta n_{s}}{Q}I_{c} \\ -|\frac{G\eta_{b} n_{s}}{Q}I_{c}|h - |\frac{G\eta n_{s}}{Q}I_{c}|sign(I_{c})M \end{bmatrix}
\end{equation}
where $U_{1}$ is the voltage of the RC circuit, which has resistance and capacitance of $R_{1}$ and $C_{1}$, in Ohms and Farads, respectively. The parameter $\eta_{b}$ is the unit-less charging/discharging efficiency of the battery, $n_{s}$ is the number of cells considered in series, $Q$ is the battery capacity in Ampere-hours (Ah), $z$ is the state of charge of the battery and $h$ is the hysteresis voltage. The hysteresis voltage dynamics contain additional parameters G and M which are tuned to match experimental results from \cite{2015Plett_BMS_Vol1}. The voltage and power of the battery can then be calculated as
\begin{equation}
    V = OCV(z) - R_{0}I_{c} - U_{1} + h
\end{equation}
\begin{equation} \label{eq:battery_power}
    P = n_{s}n_{p}VI_{c}
\end{equation}
where $OCV()$ is a lookup table that contains the open circuit voltage as a function of $z$. $R_{0}$ represents the equivalent series resistance of the cell and $n_{p}$ is the number of cells in parallel. These parameters and the model definition were sourced from \cite{2015Plett_BMS_Vol1} and a related GitHub Python repository \cite{ESCtoolbox}.

The battery storage system has both a charge and discharge capability, which is captured in the state of charge dynamics with a positive current denoting a charging state and a negative current denoting a discharge state. For the purposes of this work it is assumed that the the battery dynamics are only valid on the sets described by
\begin{equation}
    S_{I_{c}}:=\left\{I_{c}: |I_{c}|\le I_{c,max}\right\}
\end{equation}
\begin{equation}
    S_{z}:\left\{z: z - z_{min} \ge 0 \cap \ z_{max} - z \ge 0\right\}
\end{equation} 
With this in mind, the tracking goal can be formulated through the use of an error signal
\begin{equation}
    e = \frac{1}{n_{s}n_{p}}\left(P_{b} - P_{b,sp}\right)
\end{equation}
where $P_{b}$ is defined according to (\ref{eq:battery_power}). The power output of the battery with respect to current on the set described by $S_{I_{c}}$ is assumed to be strongly monotonic. The battery current control law is then formulated as a continuous representation of the first-order Newton-Raphson method, i.e.
\begin{equation}
    \dot{I}_{c} = -\left(\nabla_{I_{c}}e(I_{c})\right)^{-1}e(I_{c})
\end{equation}
The precise form of $\nabla_{I_{c}}e(I_{c})$ can be shown as
\begin{align*}
    \nabla_{I_{c}}e(I_{c}) &= \frac{\partial{e}}{\partial{I_{c}}}\\
    &= \frac{1}{n_{s}n_{p}}\left(V + \frac{\partial{V}}{\partial{I_{c}}}I_{c}\right)\\
    &= \frac{1}{n_{s}n_{p}}\left(V - R_{0}I_{c}\right)
\end{align*}
The term $-R_{0}$ (possibly unknown) is replaced with a parameter $R_{e}$ such that our control law only relies on measurements of terminal voltage $V$ and cell current $I_{c}$. We select the value of $R_{e}$ such that the denominator in the control law has the same sign as $\nabla_{I_{c}}e(I_{c})$, which is assumed positive. Thus, it is assumed that $V - R_{0}I_{c} > 0$ and a small positive or negative number $|R_{e}| << 1$ will suffice to keep the denominator positive since $V \ge V_{min} > 0$. Then the current control law becomes
\begin{equation} \label{eq:batteryCurrentControl}
    \dot{I}_{c} = -K_{I_{c}}\left(V + R_{e}I_{c})\right)^{-1}e
\end{equation}
where $K_{I_{c}}$ is a tuning gain for error convergence rate. Then, the error dynamics are written as
\begin{equation}
    \dot{e} = V\dot{I}_{c} + \dot{V}I_{c}.
\end{equation} 
Considering that the battery initially starts from rest, the hysteresis and RC voltages can be bounded due to the finite charging or discharging time of the battery by setting $I_{c} = I_{c,max}$, and noting that these are now stable linear systems, and thus, BIBO stable. Therefore, with bounded $U_{1}, \ z, \ I_{c},$ and $h$, then $\dot{V}I_{c}$ is bounded with maximum value $d_{\infty} = I_{c,max}||\dot{V}||_{\infty}$.

Next, in accordance with Definition \ref{def:ISS}, a Lyapunov function $W(e):=1/2e^{2}$ is selected which is bounded such that $\alpha_{1}(|e|) \le W(e)\le \alpha_{1}(|e|)$ where $\alpha_{1}(|e|) = \alpha_{2}(|e|)=e^{2}$. Its derivative can then be shown to be
\begin{align*}
    \dot{W}(e) &= e\dot{e} \\
    &= e\left(V\dot{I}_{c} + \dot{V}I_{c}\right)\\
    &= e\left(-K_{I_{c}}e\frac{V}{V + R_{e}I_{c}} + \dot{V}I_{c}\right) \\
    &= -Ae^{2} + eI_{c}\dot{V}\\
    &\le -Ae^{2} + ed_{\infty}\\
    &\le -Ae^{2} + \left(\frac{1}{2}e^{2} + \frac{1}{2}d_{\infty}^{2}\right)\\
    &\le -(A - \frac{1}{2})e^{2} + \frac{1}{2}d_{\infty}^{2}
\end{align*}
where $A$ is defined as
\begin{equation}
    A = K_{I_{c}}\frac{V}{V + R_{e}I_{c}}
\end{equation}
In order to establish $W(z)$ as an ISS Lyapunov function, first show that $A > 1/2$, which can be accomplished through the gain $K_{I_{c}}$ as long as $\frac{V}{V + R_{e}I_{c}} > 0$, which is always true since $V > 0$ and $R_{e}$ is designed such that $V + R_{e}I_{c}>0$. Then $\dot{W}$ can be expressed as
\begin{equation}
    \dot{W} \le (1 - 2A)W + \frac{1}{2}d_{\infty}^{2}
\end{equation}
which can be solved to yield an upper bound on the Lyapunov function as
\begin{equation}
   W(t) \le e^{(1 - 2A)t}W(0) + \frac{d_{\infty}^{2}}{2(2A - 1)}\left(1 - e^{(1-2A)t}\right).
\end{equation}
Taking the limit as $t$ approaches infinity and substituting in for $W(t)$ yields an upper bound on the error
\begin{equation}
    \lim_{t\to\infty}|e(t)| \le \frac{1}{\sqrt{2A - 1}}d_{\infty}
\end{equation}
It remains to enforce the boundedness conditions on the state of charge and current, which will be accomplished through the use of control barrier functions characterized by the safe sets below:
\begin{equation}
    b_{I_{c,min}}:=I_{c} - I_{c,min} \ge 0
\end{equation}
\begin{equation}
    b_{I_{c,max}}:=I_{c,max} - I_{c} \ge 0
\end{equation}
\begin{equation} \label{eq:h_zmin}
    b_{z_{min}}:=z - z_{min} \ge 0
\end{equation}
\begin{equation}
    b_{z_{max}}:=z_{max} - z \ge 0
\end{equation}
Similar to the approach in Section \ref{sec:windfarm} , the minimization goal of the CBF-QP will be to produce a control that is as close to possible to the control defined in (\ref{eq:batteryCurrentControl}) while maintaining forward invariance with respect to the above safe sets. An auxiliary control input $\nu$ is defined such that $\dot{I}_{c} = \nu$ and $\nu^{*} = (\ref{eq:batteryCurrentControl})$. Defining the state vector then as $\xi = \begin{bmatrix} U_{1} & z & h & I_{c} \end{bmatrix}$, the dynamics are control-affine in $\nu$ such that
\begin{equation}
    \dot{\xi} = f(\xi) + g(\xi)\nu
\end{equation}
where $f(\xi) + g(\xi)\nu$ is defined by stacking (\ref{eq:batteryDynamics}) and the control $\dot{I}_{c} = \nu$. Starting with the current bounds (using $b_{I_{c,min}}$ as an example),
\begin{align}
    &L_{f}b_{I_{c,min}} + L_{g}b_{I_{c,min}} + \alpha(b_{I_{c,min}}) \ge 0 \nonumber \\
    &\nu + c_{I_{c}}b_{I_{c,min}} \ge 0
\end{align}
where the class-$\mathcal{K}$ function for $I_{c,min}$ is linear with a scale factor $c_{I_{c}}$. The same operation is done for $b_{I_{c,max}}$, with the same constant in place for the linear class-$\mathcal{K}$ function. 

Next, for the state of charge ZCBFs, it can immediately be seen that the relative degree of these dynamics is greater than 1 (i.e. the control $\nu$ shows in the expression after differentiating $b_{z_{min}}$ more than once). Therefore, a HOCBF architecture \cite{2022Wei_HighOrderControlBarrierFunctions} as shown in Definition \ref{def:HOCBF} and Theorem \ref{thm:HOCBF} is implemented. It can be seen that this is an HOCBF with $m=2$, thus we differentiate (\ref{eq:h_zmin}) along the dynamics $\dot{\xi}$ twice to arrive at
\begin{align} \label{eq:z_psi1}
    \psi_{1}:&=\frac{\partial{b}}{\partial{\xi}}\dot{\xi} + c_{z_{min,2}}b(\xi) \ge 0\nonumber \\
    &=\frac{\eta n_{s}I_{c}}{Q} + c_{z_{min,1}}b(\xi) \ge 0 \\
    \psi_{2}:&=\frac{\partial{\psi_{1}}}{\partial{\xi}}\dot{\xi} + c_{z_{min,2}}\psi_{1} \ge 0 \nonumber \\
    &=\left(\frac{\eta n_{s}}{Q}\right)\nu + c_{z_{min,1}}\frac{\eta n_{s}I_{c}}{Q} + c_{z_{min,2}}\psi_{1} \ge 0
\end{align}
where $b(\xi)=b_{z_{min}}(\xi)$ in (\ref{eq:z_psi1}) and $c_{z_{min,1}}$, $c_{z_{min,2}}$ are constants representing the linear class-$\mathcal{K}$ functions. A similar procedure follows for $b_{z_{max}}(\xi)$. Finally, the control can be calculated at each timestep by casting this as a CBF-QP \cite{2017Ames_ControlBarrierFunctionBasedQuadraticProgramsForSafetyCriticalSystems},\cite{2014Ames_ControlBarrierFunctionBasedQPsWithApplications} 
\begin{equation} \label{eq:nu_QPCBF}
\begin{aligned}
\nu = \argmin_{\nu} \quad & \int_{0}^{T} \frac{1}{2}\left(\nu - \nu^{*}\right)^{2} dt\\
\textrm{s.t.} \quad & \nu + c_{I_{c}}b_{I_{c,min}} \ge 0 \\
\quad & -\nu + c_{I_{c}}b_{I_{c,max}} \ge 0 \\
\quad &  \psi_{2,z_{min}}(\nu,\xi) \ge 0\\
\quad &  \psi_{2,z_{max}}(\nu,\xi) \ge 0
\end{aligned}
\end{equation}

%An example is shown in Figures \ref{fig:battery_tracking} and \ref{fig:battery_barriers} to illustrate the tracking effectiveness of the control law and the utility of the control barrier functions. A sinusoidal charging and discharging set-point is provided to the battery such that it exceeds the battery capacity at its peak, and therefore the current limits. Additionally, the battery is initialized with a high (low) state of charge in the charging (discharging) such that the barriers are reached prior to the end of the simulation.

%\begin{figure}[h]
%\centering
%\includegraphics[width=0.5\textwidth]{battery_tracking.png}
%\caption{}
%\label{fig:battery_tracking}
%\end{figure}
%
%\begin{figure}[h]
%\centering
%\includegraphics[width=0.5\textwidth]{battery_barriers.png}
%\caption{}
%\label{fig:battery_barriers}
%\end{figure}

\subsection{Supervisory Control} \label{sec:superControl}

The wind-hybrid-open-controller \cite{NREL_WHOC_2025} version 0.5.1 is used here to illustrate each model's ability to perform tracking of power set-point commands with the goal of meeting a demand signal. The rule-based controller is based on heuristics and accomplishes HPP command tracking by allowing battery to contribute when there are insufficient resources available for wind and solar to meet the requested demand. It is able to reduce power when state of charge exceeds a certain threshold and stops command integration when resource saturation has occurred. This controller has also been validated on the Hercules platform \cite{NREL_HERCULES_2025} \cite{2025_ACC_WHOC_Starke}. A similar test case as shown in \cite{2025_ACC_WHOC_Starke} will be used here to motivate the use of this hybrid power plant framework.
\section{Simulation} \label{sec:simulation}

Here the models shown in Section \ref{sec:HPP_MandC} will be incorporated with the hybrid supervisory control law shown in Section \ref{sec:superControl} to illustrate the collective ability of the HPP to track a PJM RegA demand signal \cite{PJM}. The wind farm size being considered here is 160 MW and is arranged in 8 rows by 4 columns with each turbine represented by the NREL 5MW Reference turbine \cite{2007Jonkman_DynamicsModelingAndLoadsAnalysisOfAnOffshoreFloatingWindTurbine}. The rotor inertia and diameter are sourced from \cite{2007Jonkman_DynamicsModelingAndLoadsAnalysisOfAnOffshoreFloatingWindTurbine} and the wake parameters are set as $k_{w} = 0.04$ and $\Delta x = 7D_{r} \ [m]$. The parameter representing the linear class-$\mathcal{K}$ function for the wind turbine barrier function, $c_{w}$, is set to 1. The solar plant has an area $A = 1e5 \ [m^{2}]$, efficiency $\eta_{sf} = 0.5$ and is designed with a time delay, $\tau = 10 \ [sec]$ and gains $K_{p} = 2.5$ and $K_{i} = 0.2$. 

As mentioned in Section \ref{sec:batteryMandC}, the parameters are sourced from \cite{ESCtoolbox}, which is based on \cite{2015Plett_BMS_Vol1}. The parameters contained therein are scheduled with temperature, and for the purposes of this work a static temperature of \SI{25}{\degree C} is used. For the use cases here, we select a charging/discharging capacity of 40MW and an energy capacity of $E_{c} = 160 \ [MWh]$, which yields a max C-rate of 0.25C, indicating it takes 4 hours to completely charge or discharge the battery. This informs the current limits of the battery (and thus each cell) which can be determined by the multiplying the cell capacity, $Q_{c}$, by the max C-rate. The number of cells in parallel and in series are set such that $n_{s}=n_{p}=\sqrt{E_{c}/V_{c,nom}Q_{c}}$, where $V_{c,nom} = 3.3 \ [V]$ and $Q = n_{p}Q_{c}$. The state of charge limits are set as $z_{min} = 0.1$ and $z_{max} = 0.9$. The gain $K_{I_{c}}$ and the tuning parameter $c_{I_{c}}$ for the class-$\mathcal{K}$ functions related to the ZCBFs $h_{I_{c,min}}$ and $h_{I_{c,max}}$ can be tuned in tandem to set the closed loop bandwidth of the battery model. Here they are set as $K_{I_{c}} =20/n_{s}n_{p}$ and $c_{I_{c}} = 20$. The constants $c_{z_{min,1}}$, $c_{z_{min,2}}$, $c_{z_{max,1}}$, and $c_{z_{max,2}}$ are all set to 1. Finally, the static gain for the denominator in (\ref{eq:batteryCurrentControl}) is set as $R_{e} = 0$.

\begin{figure*}[h]
\centering
\includegraphics[width=1\textwidth]{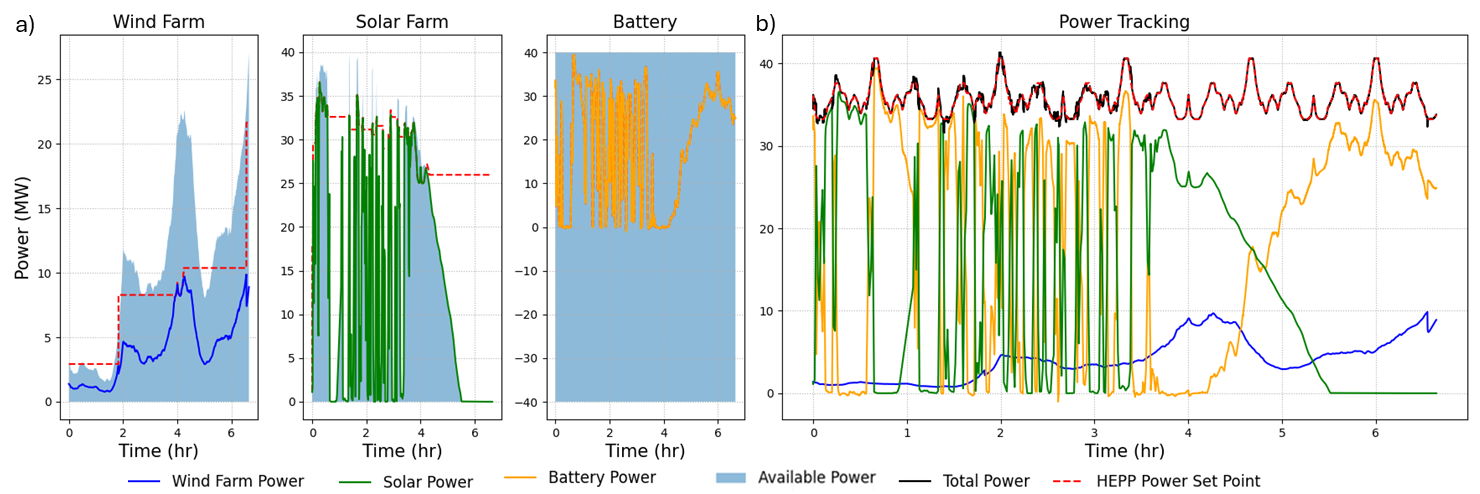}
\caption{Note that in both figures, the sign of the battery power is positive for discharging and negative for charging, which is opposite the sign convention in the dynamics but is used here for easier interpretation. In a), wind and solar resources are seen to be saturated for a majority of the simulation. Battery is able to quickly respond to allow the HPP to maintain good tracking of the demand signal. In b), tracking of the demand signal is illustrated with some difficulty when solar resource availability either ramps up or down quickly due to the irradiance quickly changing ($t = 1.5$ [hr] $\to \ t = 3.0$ [hr]). During periods when there is excess natural resource availability, the HPP is able to track the demand signal without battery contribution and the battery lies dormant.}
\label{fig:HPP_Tracking}
\end{figure*}

The results shown in Figure \ref{fig:HPP_Tracking} b) illustrate the closed loop tracking of the demand signal by the combined resources of the wind farm, solar plant, and battery which make up the HPP. There are notable times during the simulation where the tracking error of the combined HPP does not integrate to zero, which is due to the rate at which the resource availability changes and the agility of the supervisory control law to allocate set point commands to the battery to accommodate those changes. Further, Figure \ref{fig:HPP_Tracking} a) shows that the solar and wind resources are saturated often during the simulation. Specifically, for the wind farm, since the maximum available power is computed assuming the turbines are each producing power according to the incoming wind, there will always be some time varying efficiency factor offset due to the influence of wakes on the power output of the turbines. 

\section{Conclusions and Future Work}

This paper introduced an analytical control-affine Hybrid Power Plant (HPP) modeling framework consisting of a linear solar plant model, and nonlinear control affine models for both a wind farm and battery storage elements. Combined, these models were shown to track a utility demand signal using a heuristics-based supervisory control law. Further, unique control laws were developed for both the wind farm and battery storage elements that incorporated nonlinear control with ZCBFs and HOCBFs to accomplish asymptotic stability (in the case of the wind farm) and Input-to-State stability (in the case of the battery model). 
Future work will look to develop a supervisory control law to improve on some of the issues shown here (i.e. resource availability, rate limiting, command saturation, etc.) by incorporating them explicitly into the control solution through the use of predictive or optimization based control. Further, this work lacks an intelligent wind farm control law which would significantly improve its ability to reach the maximum available wind availability. Finally, time varying temperature or state of health modeling could be added to improve the battery model and further inform control decisions at both the battery and supervisory level.

\bibliographystyle{IEEEtran}
\bibliography{references}
\end{document}